\documentclass[osajnl,twocolumn,showpacs,superscriptaddress,10pt]{revtex4-1}

\usepackage{amsmath,amssymb,graphicx}

\begin{document}

\title{Designing and experimental verification of a photoacoustic flow sensor using computational fluid dynamics}

\author{Mikael Lassen}\email{Corresponding author: ml@dfm.dk}
\affiliation{Danish Fundamental Metrology, Kogle Alle 5, 2970 H{\o}rsholm,  Denmark}

\author{David Balslev-Harder}
\affiliation{Danish Fundamental Metrology, Kogle Alle 5, 2970 H{\o}rsholm,  Denmark}

\author{Anders Brusch}
\affiliation{Danish Fundamental Metrology, Kogle Alle 5, 2970 H{\o}rsholm,  Denmark}

\author{Nikola Pelevic}
\affiliation{VSL - The Dutch Metrology Institute, Thijsseweg 11, 2629 JA Delft, The Netherlands}

\author{Stefan Persijn}
\affiliation{VSL - The Dutch Metrology Institute, Thijsseweg 11, 2629 JA Delft, The Netherlands}

\author{Jan C. Petersen}
\affiliation{Danish Fundamental Metrology, Kogle Alle 5, 2970 H{\o}rsholm,  Denmark}

\begin{abstract}
A photoacoustic (PA) sensor for fast and real-time gas sensing is demonstrated. The PA sensor is a standalone system controlled by a Field-Programmable Gate Array (FPGA). The PA cell has been designed for flow noise immunity using computational fluid dynamics (CFD) analysis. The aim of the CFD analysis was to investigate and minimize the influence of the gas distribution and the flow noise on the PA signal. PA measurements were conducted at different flow rates by exciting molecular C-H stretch vibrational bands of hexane (C$_6$H$_{14}$) and decane (C$_{10}$H$_{22}$) molecules in clean air at 2950 cm$^{-1}$ (3.38 $\mu$m) with a custom made mid-infrared interband cascade laser (ICL). We observe a (1$\sigma$, standard deviation) sensitivity of 0.4 $\pm0.1$ ppb (nmol/mol) for hexane in clean air at flow rates up to 1.7 L/min, corresponding to a normalized noise equivalent absorption (NNEA) coefficient of 2.5$\times 10^{-9}$ W cm$^{-1}$ Hz$^{-1/2}$, demonstrating high sensitivity and fast real-time gas analysis. An Allan deviation analysis for decane shows that the detection limit at optimum integration time is 0.25 ppbV (nmol/mol).
\end{abstract}

\maketitle

\section{Introduction}
\label{sec:intro}  

The development of fast real-time and sensitive trace gas sensors are of increasing importance for many different applications, including environmental gas measurements, tail pipe measurements in relation to engine developments and industrial applications, portable emissions measurement systems for automobiles, biological monitoring, food quality control and for security and counter terrorism \cite{Sigrist2003,Smith2014,Hodgkinson2013,Thorpe2008, Sampaolo2016,Frey2003}. A number of different gas measurement solutions exist, including gas-chromatography, mass spectrometers, electrochemical, semiconductor sensors and laser optical absorption sensors, however they do not satisfy the growing demand for high sensitivity, high speed (real-time measurements), being non-invasive, being applicable in-situ, highly compact and cost effective. The lack of reliable sensors has therefore forced critical industries to rely on manual sampling and follow-up laboratory analysis which is labour intensive, costly, inefficient and cannot guarantee conformance of industrial and health adopted regulatory norms.

Photoacoustic spectroscopy (PAS) has the potential to close this gap. It is a direct method for monitoring the non-radiative molecular vibration relaxation modes and thus it complements techniques based on direct optical absorption spectroscopy \cite{Harren2000,Dumitras2007,Nagele2000}. Using LEDs, DFBs or quantum cascaded (interband) lasers (QC(I)Ls) as light sources PA sensors can be made both compact and still be extremely sensitive, with the ability to detect ppt (and even sub-ppt) concentrations and trace components of complex mixture of gases \cite{Bauer2014,Spagnolo2012,lassen2016OL,Zhou2017,Helman2017}. However, acoustic background noise limits the performance of the PA sensor. The noise originates from non-selective absorption in the cell windows and walls and ambient acoustic and flow noise, thus decreasing the sensitivity and accuracy of the PA measurements. Thus a fast sampling and sensitive PA sensor should be designed so that it only amplifies the PA signal from the sample under investigation and reject acoustic, flow and electrical noise sources as well as the in-phase background signals from the walls and windows. Efficient noise suppression and improved signal to noise ratio (SNR) of PA sensors has been investigated using many different designs and methods, aiming at various aspects of signal improvement, noise reduction, and ease of use \cite{Saarela2010,Saarela2011,lassen2014,Manninen2012,Dong2014,Hongpeng2017}.

Most microphone based PA sensors have relatively low resonance frequencies (0.5- 5 kHz), which makes them more sensitive to noise sources from the environment and gas flow noise. We have demonstrated that with appropriate design of the PA cell and the surrounding flow buffer zone, the coupling of flow and background noise can be reduced significantly \cite{lassen2017}. The PA sensor has been designed for flow noise immunity using advanced Computational Fluid Dynamics (CFD). We find for flow rates of up to 1.7 L/min a sensitivity of 0.4 ppb (nmol/mol) for hexane (C$_6$H$_{14}$) in clean air with an integration time of 3 seconds. This corresponds to a normalized noise equivalent absorption (NNEA) coefficient of 2.5$\times 10^{-9}$ W cm$^{-1}$ Hz$^{-1/2}$. The NNEA was derived using data from the (PNNL) quantitative Infrared Database for Gas-Phase Sensing \cite{PNNLdata}. To the best of our knowledge this is the lowest NNEA coefficient obtained for a microphone based PA sensor, when flowed with more than 1 L/min of gas \cite{Nagele2000,Saarela2010}. Besides the high sensitivity of the PA sensor, a key characteristic of the sensor system is the real-time and fast response time. The gas in the PA sensor is completely changed every 2 seconds, ensuring fast and real-time sampling. The PA sensor is not limited to molecules with C-H stretching modes, but can be tailored to measure any trace gas by simply changing the excitation wavelength (i.e. the laser source). Hexane and decane were chosen because the PA sensor aims to solve a major problem plaguing several industries that need detection of oil contaminants in compressed air \cite{lassen2017}. The PA sensor is controlled by a FPGA and is a complete standalone and self-maintaining system, enabling the construction of a commercial PA sensor.

\section{Flow simulation for optimization}

\begin{figure}[htbp]
\centering
\fbox{\includegraphics[width=\linewidth]{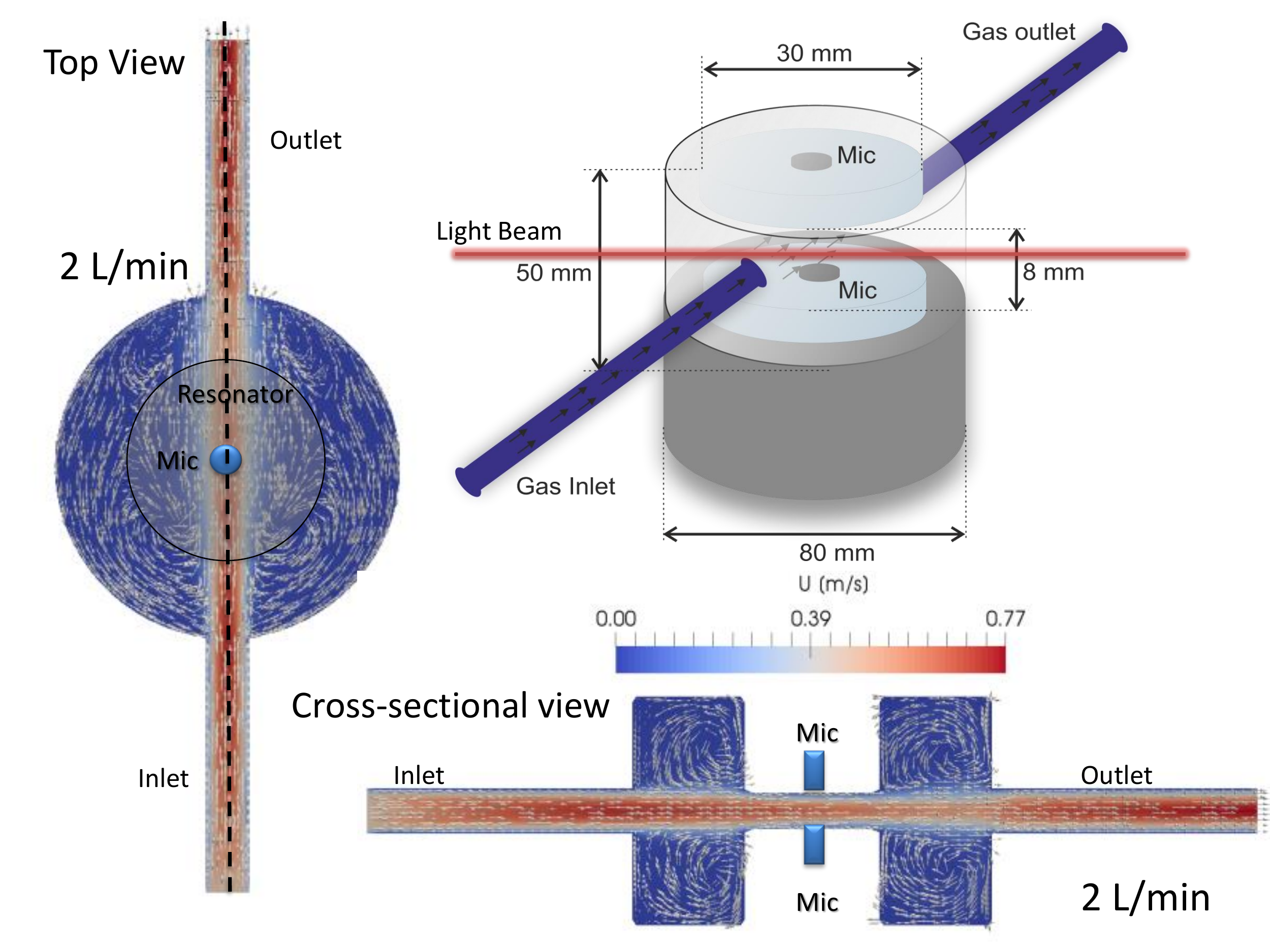}}
\caption{The flow velocity field of the gas with an inlet flow rate of the 2 L/min. The flow simulation was solved as transient and steady, laminar, incompressible flow with heat transfer. The vectors are not scaled. The flow simulation is shown from top view and cross-sectional. The dashed line in the top view illustrates the cross-sectional view. The illustration shows the design of the PA cell and dimensions.}
\label{fig1-flow}
\end{figure}

Fig.~\ref{fig1-flow} shows the schematics of the PA sensor. For fast and real-time measurements the sensor is expected to be operated at a flow of more than 1.5 L/min. This means that the PA cell with buffer zone is completely refueled 27 times per minute and the PA active area (the acoustic resonator) 116 times per minute. In order to optimize the flow and geometry of the cell, various simulations have been performed. The physical parameters that was optimized with the simulation was the position and size of inlet and outlet nuzzles and geometry and size of the buffer zone. The optimal flow was identified from the simulations of steady, laminar and incompressible flow. Fig.~\ref{fig1-flow} shows the flow velocity field of the gas within the computational domain. The fluid modeled is air with inlet flow rate of the 2 L/min. The dimensions of the nozzles are 10 mm in diameter. In order to have a continuous flow of gas through the PA cell, input nozzles and output nozzles have been attached on opposite sides of the flow cell. The acoustic resonator is enclosed by an outer cell with a 80 mm diameter. It acts as the buffer zone for the acoustic flow noise and absorption losses in the windows. The acoustic resonator is used to resonantly amplify the acoustic signal in the gas as a result of absorption in the sample. The resonator consists of two circular plates with diameter of 30 mm separated by approximately 8-10 mm. The distance was chosen experimentally, because it maximizes the PA signal, while keeping the background signal at a minimum. The plates have holes where the microphones are located. Using this PA sensor design the gas that enters the cell will flow freely through the cell without gas concentrations being accumulated in the cell, thus high recirculation volumes will be avoided. The aim of the analysis is to investigate the influence of the cell geometry and the fluid flow rate on the fluid distribution within the cell and how the flow noise is affecting the acoustics resonator. From the simulation we find that by increasing the flow speed the flow recirculation in the zones of the cell have been intensified. The intensified recirculation has negative influence on the uniform gas sampling process. At higher flow rates (more than 8 L/min) recirculation regions can give rise to longer residence time of the gas, causing wrong concentration estimates, thus gas becomes entrapped in the upper and lower regions of the buffer volume and does therefore not get flushed out. Another consequence of high flow speed is significant flow noise limiting the performance, sensitivity and reliability. However, in this study higher flow rates are not considered and the magnitude of the recirculation regions compared to the magnitude of the bulk flow can be neglected for the flow rates investigated here. The generated flow noise can be estimated using the CFD analysis and depends on the flow velocity and the tube size (cross sectional area) using the empirical equation: LN=$10 + 50 \log (v) + 10 \log (A)$, where LN is the sound power level (dB), $v$ is the gas velocity (m/s) and $A$ is the air cross sectional area (m$^2$). For 2 L/min flow this will result in a SN of -30 dB, which is comparable to a NNEA coefficient in the order of approximately $10^{-10}$ W cm$^{-1}$, equivalent to a hexane detection level of $\approx$50 ppt (pmol/mol). The results of the simulation suggest that the PA sensor could have a sensitivity in the $10^{-10}$ W cm$^{-1}$ range, this is investigated experimentally in the following.

\section{Experimental setup}

\begin{figure}[htbp]
\centering
\fbox{\includegraphics[width=\linewidth]{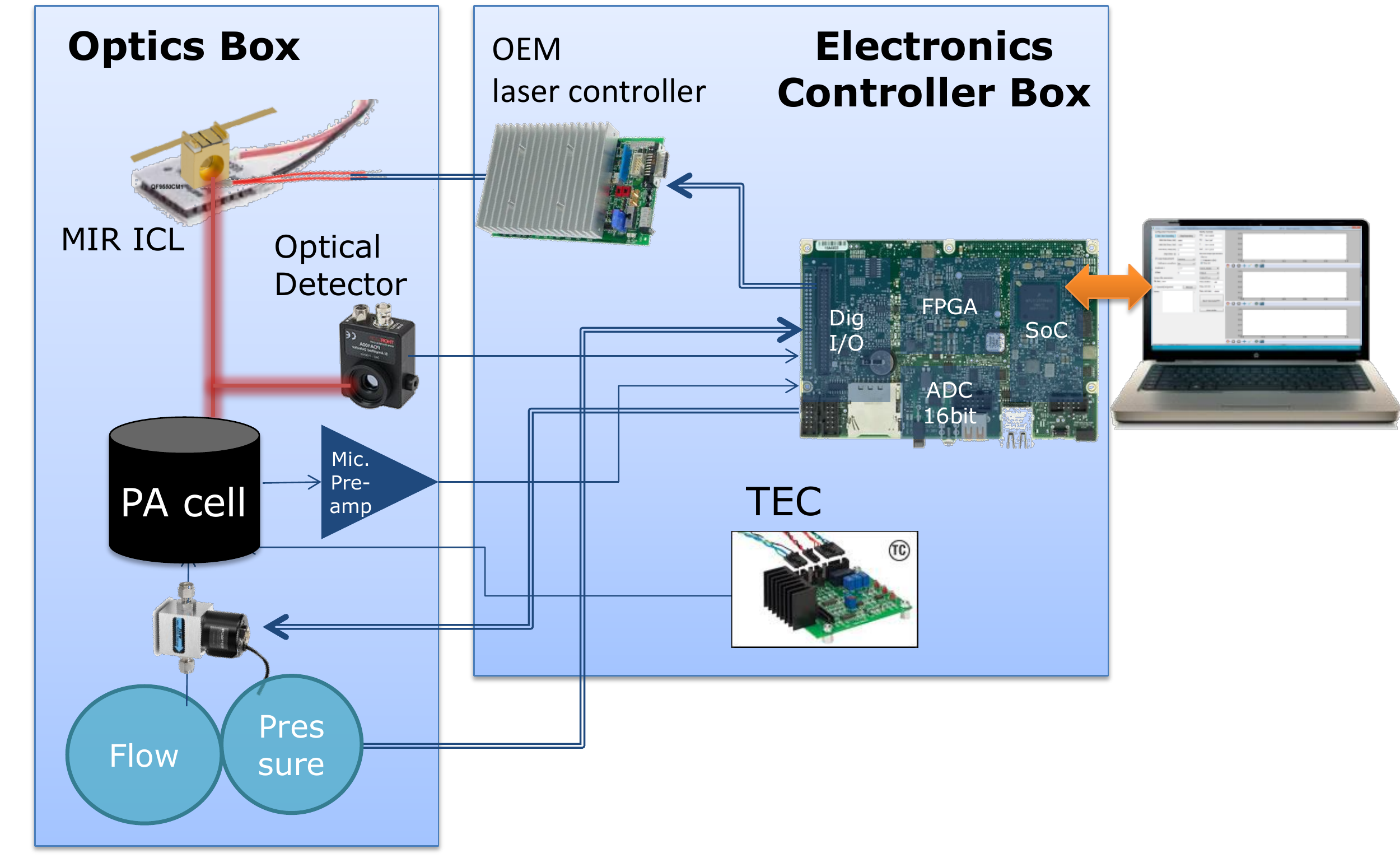}}
\caption{Schematic layout of the PA sensor based on the sb-RIO FPGA board. The PA sensor is a fully standalone system.}
\label{fig2}
\end{figure}

The schematics of the standalone real-time PA sensor for fast sampling is shown in Fig.~\ref{fig2}. The PA sensor consists of an optical box and an electronics box. The electronics box contains the power supplies, the laser driver and TEC cooler unit, the heater controller for the PA cell and the field-programmable gate array (FPGA) Single-Board RIO (sb-RIO). The two boxes are interconnected by means of cables, functionally divided to allow standard cables to be used while ensuring signal integrity. Having the power supplies located in a separate box reduces coupling of switching noise into the sensitive parts of the setup. With the multiple tasks needed to be handled for the PA sensor we found that an FPGA solution would be most suitable. We use a sb-RIO solution from National Instruments, because of the readily suitable boards and simplicity in the development tool chain where all parts of the program are collected in one project tree all based on LabView. The sb-RIO has a 400 MHz microprocessor connected to a 40 MHz Xilinx FPGA and is integrated with 16 bit AD converter with 16 multiplexing channels. The lock-in-amplifier algorithms and data logging have been implemented using the 16 bit ADC converter. The optical box (sensor head) contains the laser source mid-infrared (MIR) Interband Cascade Laser (ICL) emitting radiation at 3.38 $\mu$m with a bandwidth of approximately 40 nm and optical power of 60 mW). The modulation of the light intensity is controlled by modulating the current, where a square-wave is fed into the laser controller. The duty cycle used is always 50/50.The optics (lenses, windows and mirrors) serve to collimate the laser beam, pass it through the PA cell, the optical detectors for optical power measurements and the PA cell. The collimation (IR aspheric) lens is AR coated in the wavelength region 3 - 5 $\mu$m. A beamsplitter is inserted to tap approximately 0.5$\%$ of the laser light onto a MIR photo detector (PbSe, 1.5-4.8 $\mu$m, AC-Coupled Amplifier) for optical power measurement and normalization of the PA signal. The optical transmission through the cell is approximately 97$\%$ at 3.38 $\mu$m. The microphone pre-amplifier is located in the optical box to minimize the distance between the microphone and the input of the pre-amplifier. The optical box also contains flow and pressure controllers. The PA acoustic resonator cell has an open cell configuration, and is made by two circular plates with a radius of 15 mm and separated with a distance of 8-10 mm as illustrated in Fig.~\ref{fig1-flow}. The plates have holes where the microphones are located. The plates are made from low impurity silicon wafers to minimize absorption of scattered laser light in the acoustic resonator. The acoustics response of the PA cell is simulated using a finite element model multi-physics simulation program. The calculated eigenfrequency is 6.5 kHz. All results presented here have been conducted with 2 passes of laser beam through the PA sensor. The double pass configuration is made using a HR@3500 nm mirror to reflect the beam back into the PA cell. More technical details about the PA sensor can be found in Ref.~\cite{lassen2017}.

\section{Results}

\begin{figure}[htbp]

\centering
\fbox{\includegraphics[width=\linewidth]{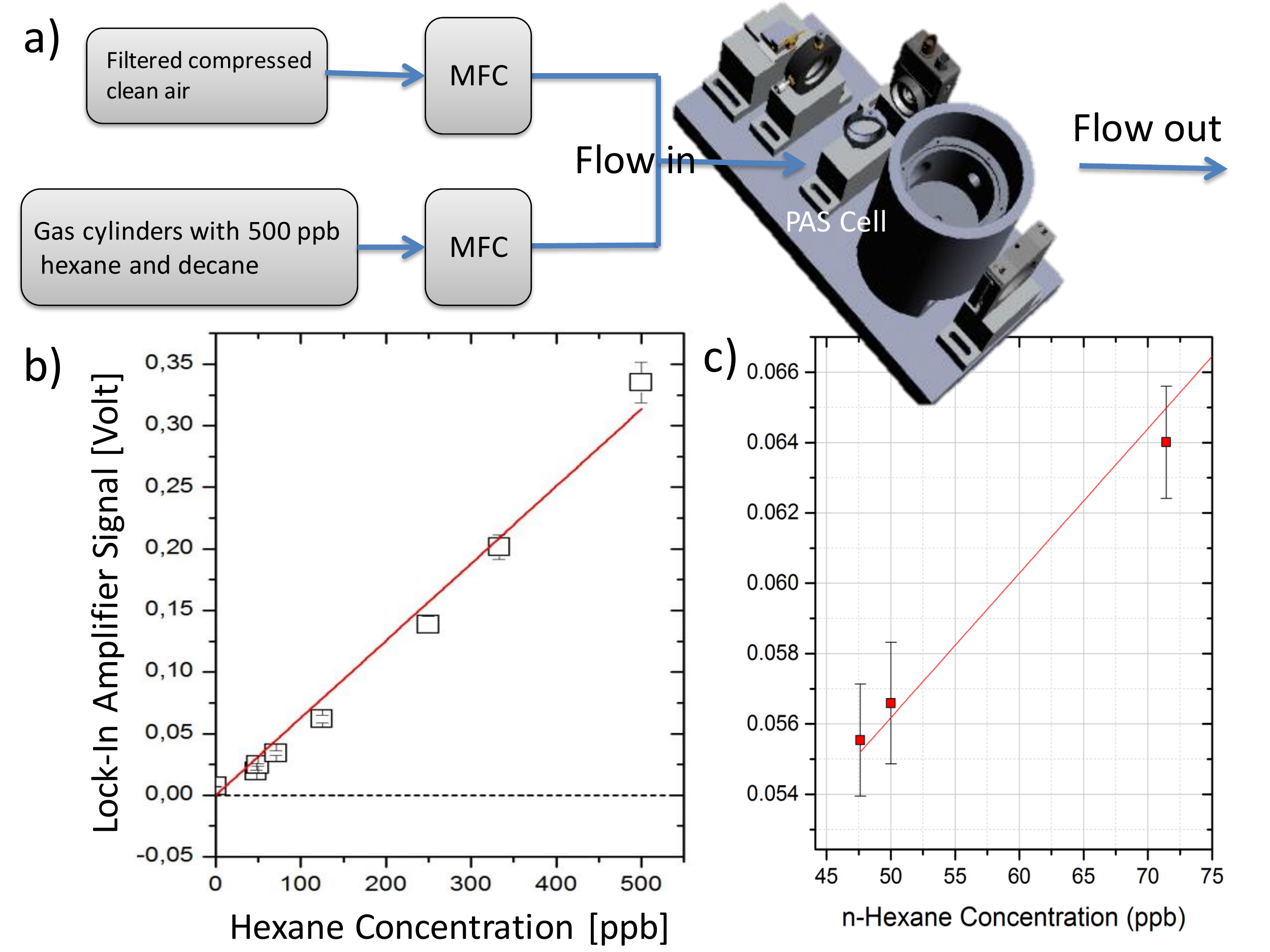}}
\caption{a) Diagram of the dilution system for linearity test of the PA sensor. MFC: Mass Flow Controller. b) Relation between concentrations based on the dilution system as function of the lock-in-amplifier signal. The linear regression is R$^2 \approx 0.99$, which indicates that the PA sensor has a high level of linear response. c) Close up on the three measured concentrations at 47.6 ppb, 50 ppb and 71.1 ppb.}
\label{fig3}
\end{figure}

The flow experiments have been conducted by purging the cell with pure air mixed with a 500 ppb concentration of hexane or decane. The reference gas used was prepared at VSL (The Dutch metrology institute) and used as input to the dilution system as shown in the diagram of Fig.~\ref{fig3}a. Two Mass Flow Controllers (MFC, Brooks) were used for diluting the 500 ppb hexane and decane reference gases with clean compressed air (Parker Balston purge gas generator). The flow of the reference gas was set to 0-100 mL/min, while the dilution flow was changed within the range of 0-2 L/min.

\subsection{Transfer function for the lock-in signal to the gas concentration}

To establish a transfer function from the lock-in signal to concentrations the average lock-in-signal at each of the dilution settings was calculated. The measurements for the dilution of the gas is shown in Fig.~\ref{fig3}b. At each dilution step measurements were performed for 5 to 15 minutes. The result of the measurement during this period was used to generate a mean value and standard deviation, thus a linearity test was performed, as shown in Fig.~\ref{fig3}b. The hexane concentrations have an uncertainty of less than 3$\%$ on each measurement. A linear fit between the average lock-in-amplifier signal and the expected concentration from the dilution system provided the relation. The linear regression is R$^2 \approx 0.99$. The PA sensor is able to detect the 2.4 ppb concentration change as a change from 50 ppb to 47.6 ppb from the dilution system, the lock-in-amplifier actually detects this change as change from 46.4$\pm$3.9 to 43.6$\pm$4.0 (n=1000, $\pm$1$\sigma$, each). The data rate is 10 Hz and the precision average to 0.4 ppb (NNEA coefficient of 2.5$\times 10^{-9}$ W cm$^{-1}$ Hz$^{1/2}$). The measurements were conducted with different flow speeds ranging from 0.15 L/min to 1.2 L/min. Thus the linearity test also confirms that the PA sensor is immune to flow noise in this flow range.

\begin{figure}[htbp]
\centering
\fbox{\includegraphics[width=\linewidth]{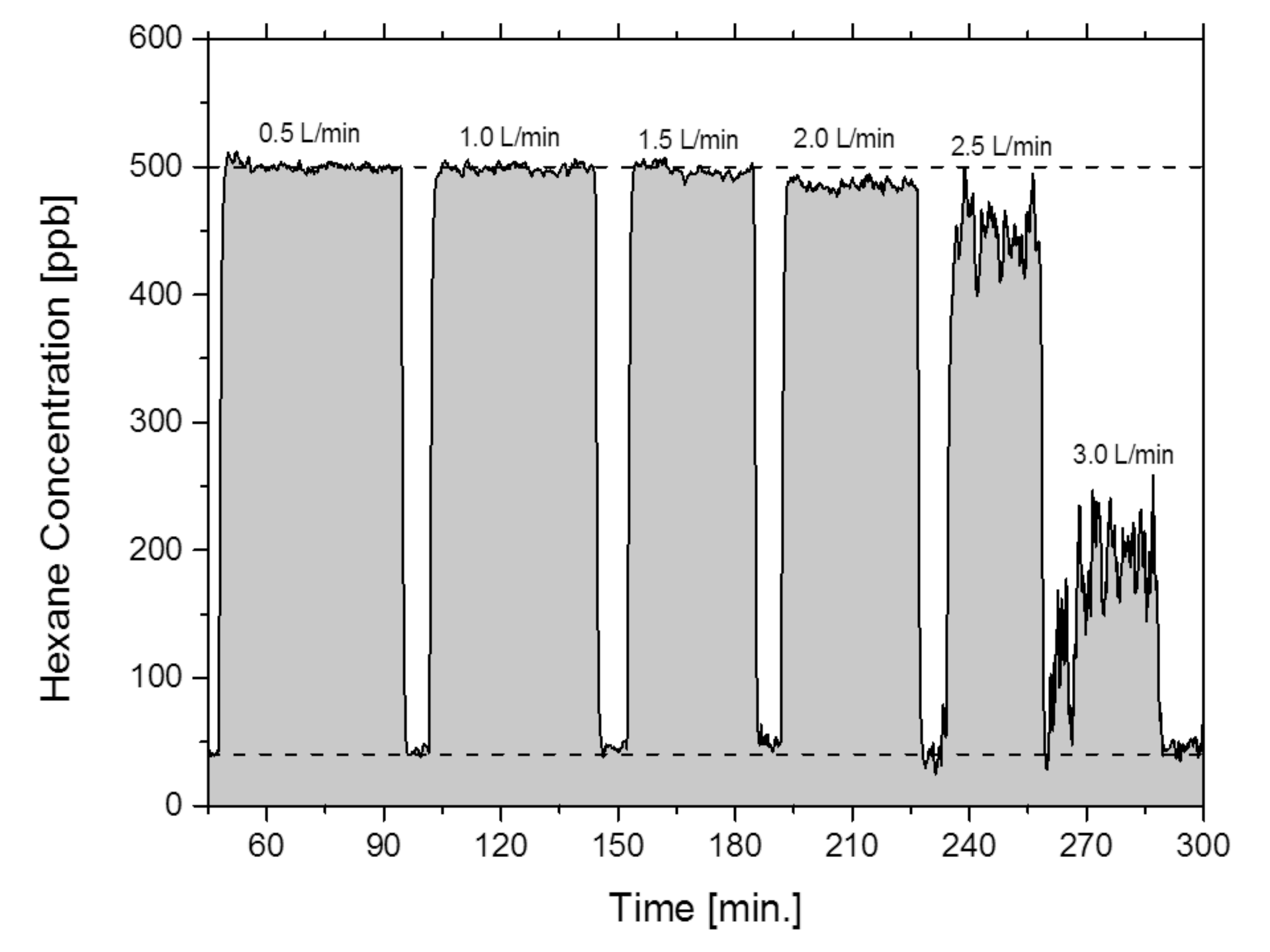}}
\caption{Sensitivity of the PA sensor as function of flow speed for 3 second integration. Response time and the reproducibility of the PA sensor for different flow speeds.}
\label{fig4}
\end{figure}

Figure \ref{fig4} shows an example where the PA cell is purged with a flow of up to 3 L/min and it is observed that even at relatively high flow rates (up to 2 L/min), the PAS cell has sufficient SNR to resolve concentration below 1 ppb of hexane (with a 3 second integration time). From Fig.~\ref{fig4} we observe a (1$\sigma$, standard deviation) sensitivity of 0.4 $\pm0.1$ ppb (nmol/mol) for hexane in clean air. From Fig.~\ref{fig4} we see that for flow speed of 2 L/min that the reproducibility is within 2.6$\%$, however the 1$\sigma$ standard deviation is not effected, thus we believe that the slightly lower level is not due to the flow noise but probably the general stability of the PA sensor. This also demonstrates that the sensor can be flowed with up to 2 L/min without a decrease in sensitivity. The response time was tested by flowing the PA sensor with 500$\pm$0.5 ppb hexane followed by a diluting the hexane with pure air to 50$\pm$4.0 ppb. Figure~\ref{fig4} shows that the reproducibility of the PA sensor is better than 1$\%$ from peak to peak for flow up to 1.5 L/min and the response time is around 10-15 seconds for reaching the full level of 500 ppb again for all flow speeds. Thus, the response time is limited by the lock-in amplifier integration time, which was 3 seconds for these measurements. Changing the integration time will however not only influence the response time but also the sensitivity of the sensor, thus low integration time will give a faster response and a lower concentration sensitivity. From this we conclude that the time response of the PA sensor will not be the limiting factor for real life practical applications.

\subsection{Allan deviation}

\begin{figure}[htbp]
\centering
\fbox{\includegraphics[width=\linewidth]{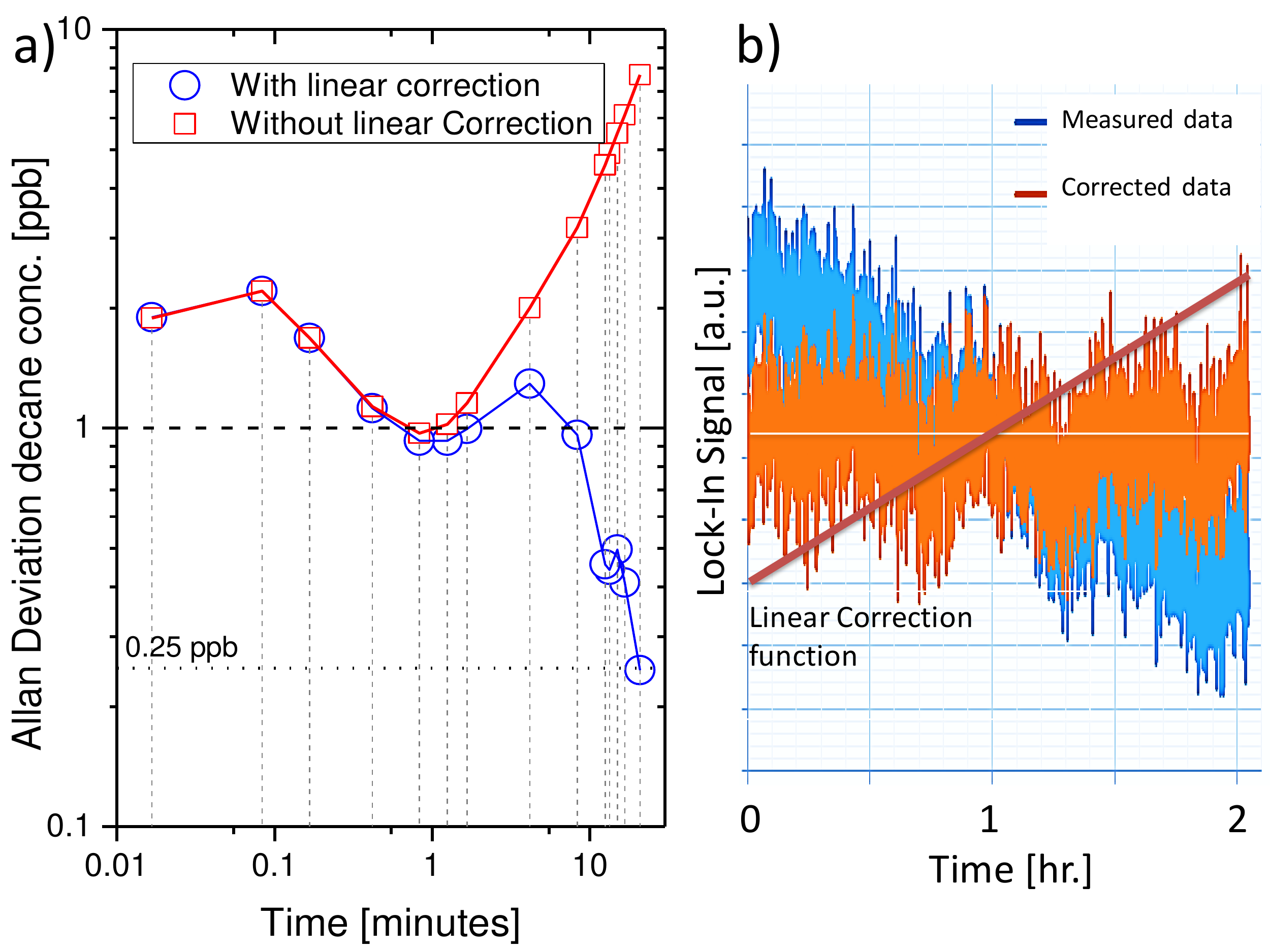}}
\caption{a) Allan Deviation of a 2 hours’ measurement on 500 ppb decane. Red curve without linear drift correction. Blue curve with linear drift correction. b) Time trace of 2 hours of measurement on 500 ppb decane with and without the linear correction. The red curve illustrates the applied simple linear correction.}
\label{fig5}
\end{figure}

The PA sensor was therefore tested for stability by performing long measurements by flowing continuously 1.7 L/min of 500 ppb decane concentration through the PA cell. The corresponding Allan deviation is shown in Fig.~\ref{fig5}a. From the figure can be seen that at very short observation time the Allan deviation is high due to noise. At longer integration times it decreases because the noise averages out. In this measurement (red curve) the Allan deviation shows instability of the system on time orders greater than 1 minute. It is seen that the stability up to 1-minute averaging lies within 1 ppb. At still longer integrations times the Allan deviation starts increasing again, suggesting that the PA sensor is drifting due to temperature changes or misalignment or other factors that may affect the estimate of the concentration. For example, the ICL is affected by a thermal drift and can therefore be eliminated by normalizing the PA signal to the optical power or by applying a linear correction to the measurement afterwards, which minimizes the STD and makes the mean become 500 ppb. The correction of the data is shown in Fig.~\ref{fig5}b applying the linear correction provides an estimate for the performance if this drift is eliminated. The blue curve in Fig.~\ref{fig5}a depicts the applied drift correction term and the Allan deviation shows that averaging keeps improving now up to 20 minutes of averaging which results in a sensitivity of approximately 0.25$\pm0.1$ ppb. We find that the absolute sensitivity for decane is approximately 1.5 times larger than the hexane response which is in good agreement with the normalized absorption as a function of carbon number given by Ref. \cite{PNNLdata}

\section{Conclusion}

We have demonstrated that even with a gas flow of more than 1.7 L/min the PA sensor can be used without a significant increase of the acoustic background noise. It was demonstrated that at relatively high flow rates the PA sensor has sufficient SNR to resolve concentration of hexane and decane below 1 ppb. Explicit we find for flow rates of up to 1.7 L/min a sensitivity of 0.4 ppb (nmol/mol) for hexane in clean air with an integration time of 3 seconds. This corresponds to a NNEA coefficient of 2.5$\times 10^{-9}$ W cm$^{-1}$ Hz$^{-1/2}$. To the best of our knowledge this is the lowest NNEA coefficient obtained for a microphone based PA sensor, when flowed with almost than 2 L/min. The PA sensor is not only limited to detect molecules in the mid-infrared, but is suitable for various practical sensor applications in the ultraviolet (UV) to the mid-infrared wavelength region simply by changing the light source. We believe that the presented PA sensor system is a novel and valuable instrument to implement for critical and continuous monitoring of trace gases. The PA sensor is controlled by FPGA and is a complete standalone system, thus it fulfils the requirements for industrial and microbial monitoring. We believe that the PA sensor has the potential to reduce the costs considerably associated with quality and safety measurements of various environmental, clinical and industrial processes.

\section*{Funding} We acknowledge the financial support from EUREKA (Eurostars program: E10132 - PASOCA) and the Danish Agency for Science, Technology and Innovation. The European Unions Seventh Framework Programme (FP7) managed by REA 8211 under Grant Agreement N. 286106. 

\section*{Acknowledgement}
We would like to thank Poul Jessen and Søren Laungaard from PAJ Group (paj@paj.dk) for fruitful discussions and collaboration.

\end{document}